# Causality in cardiorespiratory signals in pediatric cardiac patients

Maciej Rosoł[1], Jakub S. Gąsior[2], Iwona Walecka[2], Bożena Werner[2], Gerard Cybulski[1], Marcel Młyńczak[1]

*Abstract*— Four different Granger causality-based methods - one linear and three nonlinear (Granger Causality, Kernel Granger Causality, large-scale Nonlinear Granger Causality, and Neural Network Granger Causality) were used for assessment and causal-based quantification of the respiratory sinus arrythmia (RSA) in the group of pediatric cardiac patients, based on the single-lead ECG and impedance pneumography signals (the latter as the tidal volume curve equivalent). Each method was able to detect the dependency (in terms of causal inference) between respiratory and cardiac signals. The correlations between quantified RSA and the demographic parameters were also studied, but the results differ for each method.

*Clinical relevance*— The presented methods (among which NNGC seems to be the most valid) allow for quantification of RSA and study of dependency between tidal volume and RR intervals, which may help to better understand association between respiratory and cardiovascular systems in different populations.

## I. INTRODUCTION

Although the circulatory and respiratory systems serve two distinct functions, their activity is deeply interconnected and interrelated. Respiratory sinus arrhythmia (RSA), a phenomenon whereby RR intervals shorten during inhalations and lengthen during exhalations, is one example of such interactions [1]. This physiological phenomenon is rooted in many mechanisms, among which is the activity of the sympathetic and parasympathetic nervous systems [2,3]. Heart rate variability (HRV) parameters reflect the fluctuation of RR intervals and have wide applicability in medicine and sports as a means of assessing autonomous nervous system activity [4,5]. HRV parameters can be calculated from ECG or from photoplethysmography signal [6,7].

RSA can be used as a biomarker reflecting conditions like diabetes, sleep apnea, cardiomyopathy, anxiety disorders or heart failure [8-10]. Approaches based on the Granger causality concept are among the viable methods of RSA assessment and quantification [11]. In a particular Granger causality analysis, two models are used. The first one predicts the values of X signal (i.e., the tachogram/RR intervals signal) based on the past values of this signal, while the second model tries to fit the same value but based on the past values of X and Y signals (i.e., both tachogram and respiratory signal). Y is said to cause X (denoted as Y→X) if the prediction error of the second model is statistically significantly smaller than the prediction error of the first model. The measure of RSA would depend on the improvement in the prediction accuracy after incorporating the Y signal (e.g., respiratory signal) for the prediction of X signal (in this case the tachogram). Causality can be assessed and quantified from a respiratory signal to a tachogram, but also in the opposite direction.

The aim of this paper is to present and compare the results from different causality-based methods (one linear and three nonlinear) used in analysis of short-term RSA phenomenon among a group of pediatric cardiac patients.

## II. METHODS

### A. Conducted study

This study analyzed data recorded among 20 patients (a demographic breakdown is presented in Table 1). In order to be included in the study, patients had to be between 7 and 15 years old and had to have a current diagnosis of cardiac disease (as well as a signed consent form). Patients were excluded if they:

- presented symptoms of infection,
- had been diagnosed with disorders potentially affecting the autonomous nervous system, or
- were unable to perform physical activity.

TABLE I.   THE DEMOGRAPHIC DETAILS OF THE STUDY GROUP (AS MEAN ± STANDARD DEVIATION). THE NUMBER OF PARTICIPANTS PER SEX IS GIVEN IN THE BRACKETS.

| *Demographic data* | | | |
|---|---|---|---|
| | *All* | *Boys (12)* | *Girls (8)* |
| Age [years] | 12.9 ± 3.5 | 14.3 ± 3.3 | 10.8 ± 2.6 |
| Weight [kg] | 58.8 ± 23.2 | 65.2 ± 22.9 | 46.0 ± 18.1 |
| Hight [cm] | 160.1 ± 17.2 | 166.1 ± 16.3 | 151.0 ± 16.3 |
| BMI [kg/m$^2$] | 21.5 ± 5.0 | 22.7 ± 5.1 | 19.3 ± 4.1 |

The study was approved by the Ethics Committee of the Medical University of Warsaw (permission: KB/70/2021).

In this study the cardiac activity was registered through single-lead ECG, while the respiratory signal (tidal volume equivalent) was obtained from impedance pneumography (IP). The latter measurement was obtained using a Pneumonitor 2 [12]. Tetrapolar configuration was used for IP recordings with sinusoidal application current, which amplitude was up to 1 mA, and frequency was equal to 100 kHz. The electrodes were placed according to [13], with the receiving electrodes placed on a midaxillary line around 5th and 6th rib and the application electrodes attached to the proximal side of the arm at the level of receiving electrodes. Both signals had sampling frequency of 250Hz. Standard Holter-type Ag/AgCl electrodes were used.

[1]M. Rosoł, M. Młyńczak and G. Cybulski are with the Warsaw University of Technology, Faculty of Mechatronics, Institute of Metrology and Biomedical Engineering, 8 Boboli Street, 02-525 Warsaw, Poland. maciej.rosol.dokt@pw.edu.pl

[2]J.S. Gąsior, I. Walecka and B. Werner are with the Medical University of Warsaw, Department of Pediatric Cardiology and General Pediatrics, Żwirki i Wigury 63a, 02-091 Warsaw, Poland

All measurements were conducted in a diagnostic room in the Department of Pediatric Cardiology and General Pediatrics of the Medical University of Warsaw. The study protocol consisted of:
- obtaining a signed examination consent form from the patient's parent or legal guardian,
- presenting the parent/legal guardian with the conditions for the patient's participation in the study,
- presenting the testing apparatus and carrying out the process of getting the patient acclimated to the test conditions,
- commencing the recording at rest until rest values were reached,
- acquiring signals for analysis over the span of 5 minutes.

*B. Methods used*

The RR signal was obtained by detecting the R peaks in ECG using Stationary Wavelet Transform [14,15] and then performing cubic interpolation. The IP signal was filtrated using high-pass and low-pass Butterworth filters with cutoff frequencies equal to 0.1 Hz and 30 Hz, respectively. Based on [16], it was assumed that by using the aforementioned electrode configuration, the linear fitting provides the best compatibility between IP and pneumotachometry so that the IP signal was treated as relative tidal volume equivalent (TV). Both RR and TV signals were down-sampled to 25 Hz to obtain the compromise between computational complexity reduction and time resolution preservation (for analytical methods utilized).

For the assessment and quantification of the RSA the following methods were used:
- Traditional Granger causality analysis (GC) [17],
- Kernel Granger Causality (KGC) [18],
- Large scale Nonlinear Granger Causality (lsNGC) [19], and
- *nonlincausality* Python package using neural networks for prediction in term of Granger causality (NNGC) [20].

Moreover, the change of the causality relationship between the signals over time was studied using *nonlincausality* approach with windows $w1$ and $w2$, equal to 25 (1 second) and 1 (40 milliseconds) samples, respectively. The temporal orders between signals were also studied with the *tempord* package [21] using linear-based LM method with 0.9 threshold and time shift between -2 and 2 seconds.

All of the above-mentioned methods were used in quantifying RSA for each patient; subsequently, because all the methods used allow for statistical inference, the statistical significance of these findings was assessed. Analysis of causality was performed for lag equal to 1 second for each method. In the case of the KGC method, the chosen kernel was gaussian and the order parameter was 25. For lsNGC, the $c_f$ and $c_g$ parameters, which corresponds to the number of hidden neurons in the Generalized Radial Basis Functions neural network were set to 25. The architecture of neural networks used in NNGC consisted of 2 hidden layers with 20 neurons with the ReLU activation function in hidden layers and linear activation function in the output layer. Networks were trained for 100 epochs with learning rate equal to 1e-4 and 1e-5 for the first and second half of the learning process, respectively.

Dependencies of causality values in both directions were also studied in light of demographic data. Correlation between causality metrics and the patient's age and body mass index (BMI), as well as between measures of causality in different directions was calculated using Spearman's correlation coefficient. Mann-Whitney U rank test was used to check whether the patients' sex accounted for any differences in causality measures.

The results obtained from temporal orders analysis were also compared to the results from the group of 10 elite athletes and 10 healthy students presented in [21].

The significance level was assumed to be 0.05. All calculations were run using Python 3.7 (calculation of GC, lsNGC and NNGC), R version 3.6.1 (temporal orders) and MATLAB 2019b (KGC).

## III. RESULTS

*A. Causality analysis*

An example of the obtained tachogram and respiratory signal is presented in Fig. 1.

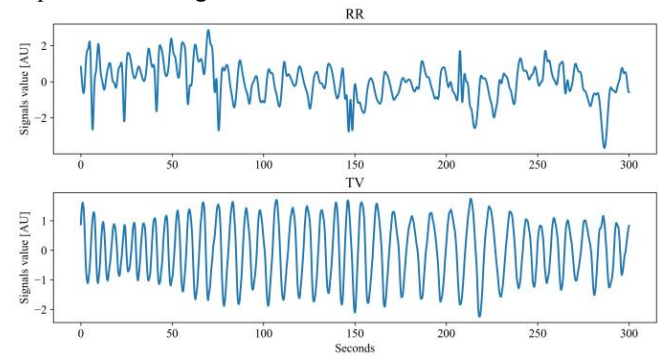

Figure 1. Sample plots of tachogram and respiratory signal from the patient #8.

For each patient the causality between both signals was quantified using each method and in both directions, and those values were plotted in Fig. 2. For twelve patients the causality from TV to RR was higher than from RR to TV, and for eight RR→TV causality was higher in case of GC and NNGC. For lsNGC method, the proportions were the opposite as eight children obtained higher causality values for the direction from TV to RR and twelve of them - from RR to TV.

Causality between signals was statistically significant in both directions and regardless of the method used, an exception being the TV→RR causality according to lsNGC in patient #4.

The results of the correlation analysis are presented in Table 2. Irrespective of the method used, there was no statistically significant correlation between age and either direction of causality. There was a significant negative correlation

(p-value < 0.05) between BMI and the TV→RR causality value calculated using NNGC (Spearman's rho equal to -0.46), and a positive one between BMI and RR→TV calculated using lsNGC (Spearman's rho equal to 0.52). For lnNGC and GC approaches, there were also significant correlations (p-values < 0.01 and < 0.05, respectively) between TV→RR and RR→TV with Spearman's rho equal to 0.59 and 0.45, respectively. There was a statistically significant difference in causality for boys and girls for lsNGC and GC (p-values < 0.05).

*B. Temporal causality*

Changes of causality over time calculated with *nonlincausality* package (NNGC approach) were plotted for each patient in both directions (TV→RR, and RR→TV). For each patient, the signal of change in causality over time was highly variable for both directions of causality, regardless of the patient's age, sex, or BMI. A sample plot of this change can be seen in Fig. 3. The temporal orders were also visualized for each patient and the example plot is presented in Fig. 4.

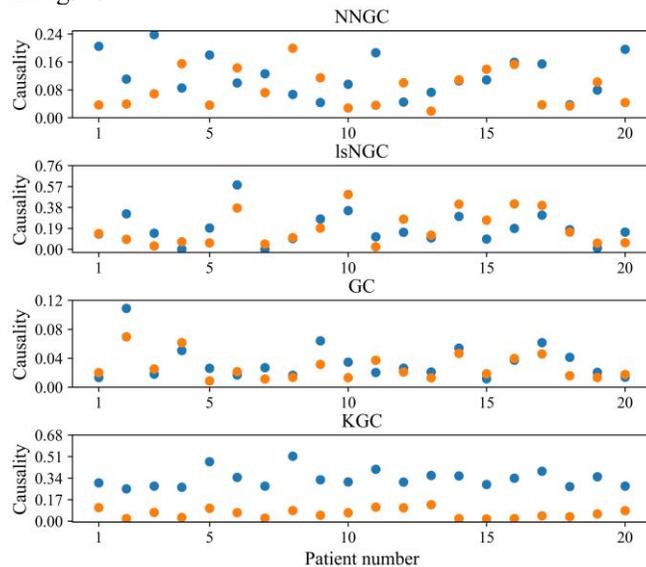

Figure 2. Causality values assessed with different methods for each individual patient. The blue dots are representing the quantified causality values from TV to RR, while orange ones stand for causality from RR to TV.

TABLE II.  SPEARMAN'S CORRELATION COEFFICIENTS WITH P-VALUES IN BRACKETS BETWEEN CAUSALITY AND AGE OR BMI AND BETWEEEN TWO DIRECTIONS OF CAUSALITY FOR DIFFERENT METHODS.

|  | *Age* | | *BMI* | | *Causalities* |
| --- | --- | --- | --- | --- | --- |
|  | *TV→RR* | *RR→TV* | *TV→RR* | *RR→TV* |  |
| GC | 0.30 (0.20) | 0.16 (0.49) | 0.25 (0.30) | 0.12 (0.63) | 0.45 (<0.05) |
| KGC | -0.38 (0.10) | -0.30 (0.21) | -0.31 (0.20) | -0.33 (0.17) | 0.40 (0.08) |
| lsNGC | 0.17 (0.47) | 0.13 (0.60) | 0.15 (0.53) | 0.52 (<0.05) | 0.59 (<0.01) |
| NNGC | -0.35 (0.14) | 0.20 (0.39) | -0.46 (<0.05) | 0.04 (0.88) | -0.19 (0.43) |

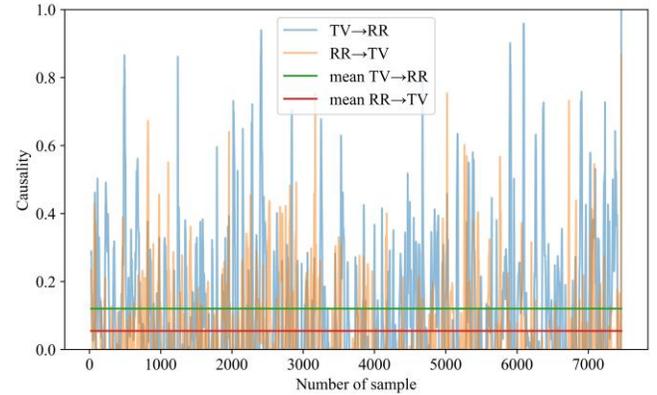

Figure 3. Causality changes over time for RR→TV and TV→RR for patient #10.

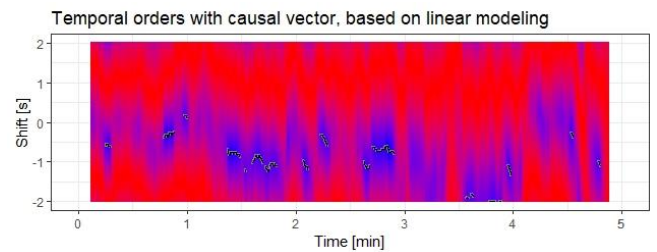

Figure 4. Sample temporal order plot for patient #8.

In line with expectations, the causal vectors obtained among cardiological patients were much smaller and less continuous than those collected in studies with subject populations of elite athletes and healthy students (presented in [21]). The average causal vector was -72 ± 668 milliseconds (mean ± standard deviation). For six of the twenty patients there was no causal vector at all.

## IV. DISCUSSION

All of the used methods were able to detect a dependency between RR and TV in at least one direction. The KGC method stands out as the only one which in every case indicated a higher causal relationship from TV to RR. Meanwhile, the GC and NNGC methods detected TV→RR as higher more frequently, while lsNGC found more cases with higher causality from RR to TV. The negative correlation between BMI and causality derived from NNGC data seems to be in line with the findings of *Mazurak et. al.* [22], which demonstrated that decreasing a patient's body mass would result in increased HRV measures. Given that the directions of the causalities for GC and lsNGC are intercorrelated it is questionable whether using both measures would be useful in case of those methods. When it comes to NNGC and KGC, the usage of both directions of causality measures may lead to new and additional information about patients.

Based on a plot of causality figures for each patient and method, it is apparent that each method interprets the two directions of causality differently. Because each one is capable of detecting causality between the analyzed signals, they could all potentially be useful, as each might provide different diagnostic information. That said, the NNGC approach appears to be the most valid as it indicates a correlation of RSA with BMI while the two directions of causality measurements are not intercorrelated.

It is also to be emphasized that the detected causality from RR to TV does not indicate that the work of the cardiovascular system is driving the work of the respiratory system, but it provides additional information regarding the interdependencies between signals from those systems. Moreover, due to the complexity of the relationships between those systems, the usage of nonlinear modeling methods seems to be a more appropriate choice.

Furthermore, it seems that the analysis of temporal orders may be a valuable tool capable of distinguishing between different group of patients. Although further studies on the matter are undoubtedly needed, analysis of causality's change over time might prove to be a source of additional information, as it can help visualize relationships between various signals and causality measures.

The foremost limitation of this study lies in its relatively small study group as well as in the homogeneity of this group. Because the study sample lacked a control group of healthy subjects, the utility of the calculated causality values as a biomarker for cardiac autonomic functions abnormalities could not be determined. Therefore, further studies (ones involving larger and more varied sample populations) will have to be conducted in order to ascertain the possible utility of each method and each direction of causality in terms of quantification of RSA. The main limitation of the mentioned techniques is their dependance of the choice of input, which might have an impact on the final result.

## V. CONCLUSION

RSA was detected and quantified for each patient using four different causality-based methods. There appears to be no correlation between the results and the patient's age, but some correlation with BMI (in the case of NNGC and lsNGC) and with sex (in the case of GC and lsNGC) may be demonstrable. Further studies on a larger group including healthy subjects are needed to assess the utility of each method in terms of RSA.